\documentclass[preprint,aps,prb,12pt]{revtex4}
\usepackage{amsfonts}
\usepackage{graphicx}
\usepackage{amsmath}
\usepackage{natbib}
\usepackage{epsfig}
\usepackage{dcolumn}

\begin{document}

\title{Suppression of telegraph noise in a CPP spin valve
 by an oscillating spin 
torque: Numerical study}

\author{A. Rebei}
\affiliation{  Seagate Research Center, Pittsburgh, Pennsylvania 15222,USA }
\email{arebei@mailaps.org}

\begin{abstract}
The phenomenon of stochastic resonance (SR) has been mainly studied in 
one-dimensional systems with additive noise. We show that in higher 
dimensional systems
and in the presence of multiplicative  noise, a 
 non-linear magnetic system  
with a strongly periodic current can show behavior similar to that
of SR but only  
 for frequencies below the ferromagnetic resonance (FMR) 
frequency of the system. \ Such a phenomena can provide
 an effective way 
to suppress low frequency noise in spin valve magnetic sensors.
\end{abstract}

\maketitle

\bigskip

\bigskip

\begin{figure}[htbp] 
\centering{\resizebox{4cm}{!}{\includegraphics{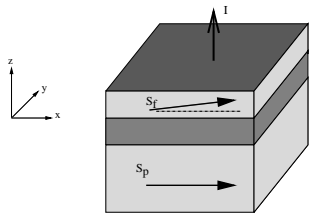}}}
\caption{The tri-layer geometry of the spin valve described in this work. The bottom magnetic layer is supposed to be very thick and is pinned along 
the x-axis. The top layer is also magnetic but free. Both layers are separated by a thin $0.8 \; nm$ normal conductor and traversed by an ac and a dc current as shown. }
\label{geometry}
\end{figure}

\begin{figure}[htbp]
\centering{\resizebox{4cm}{!}{\includegraphics{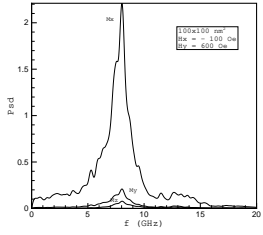}}}
 \caption{Spectral densities (in arbitrary units) at zero current of the different components of the magnetization. \ The FMR peak of the system is around $7.0 \; GHz$ in the absence of a spin torque.  }
\label{ZeroCurrent}
\end{figure}

\begin{figure}[htbp]
\centering{\resizebox{4cm}{!}{\includegraphics{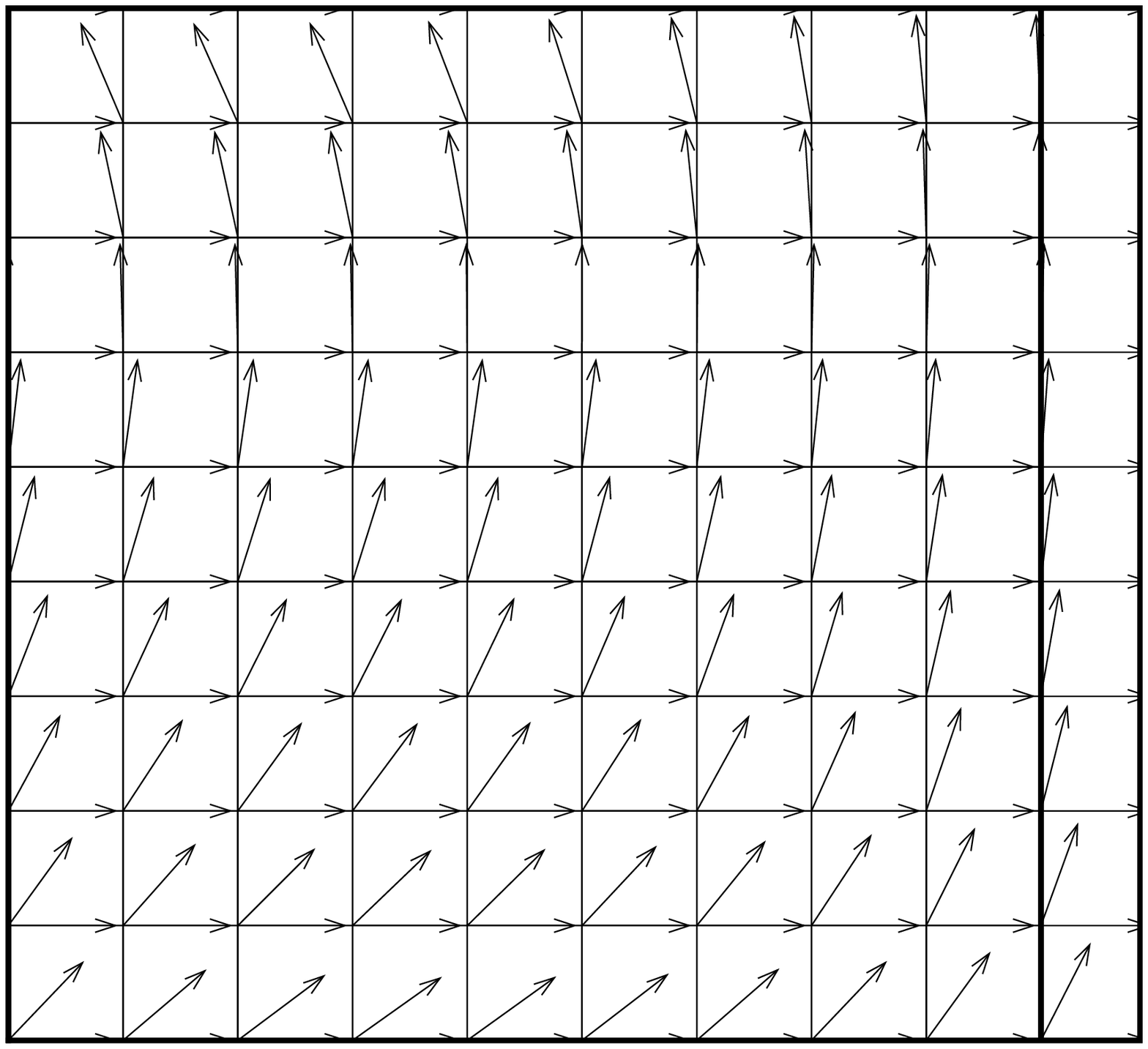}}}
\centering{\resizebox{4cm}{!}{\includegraphics{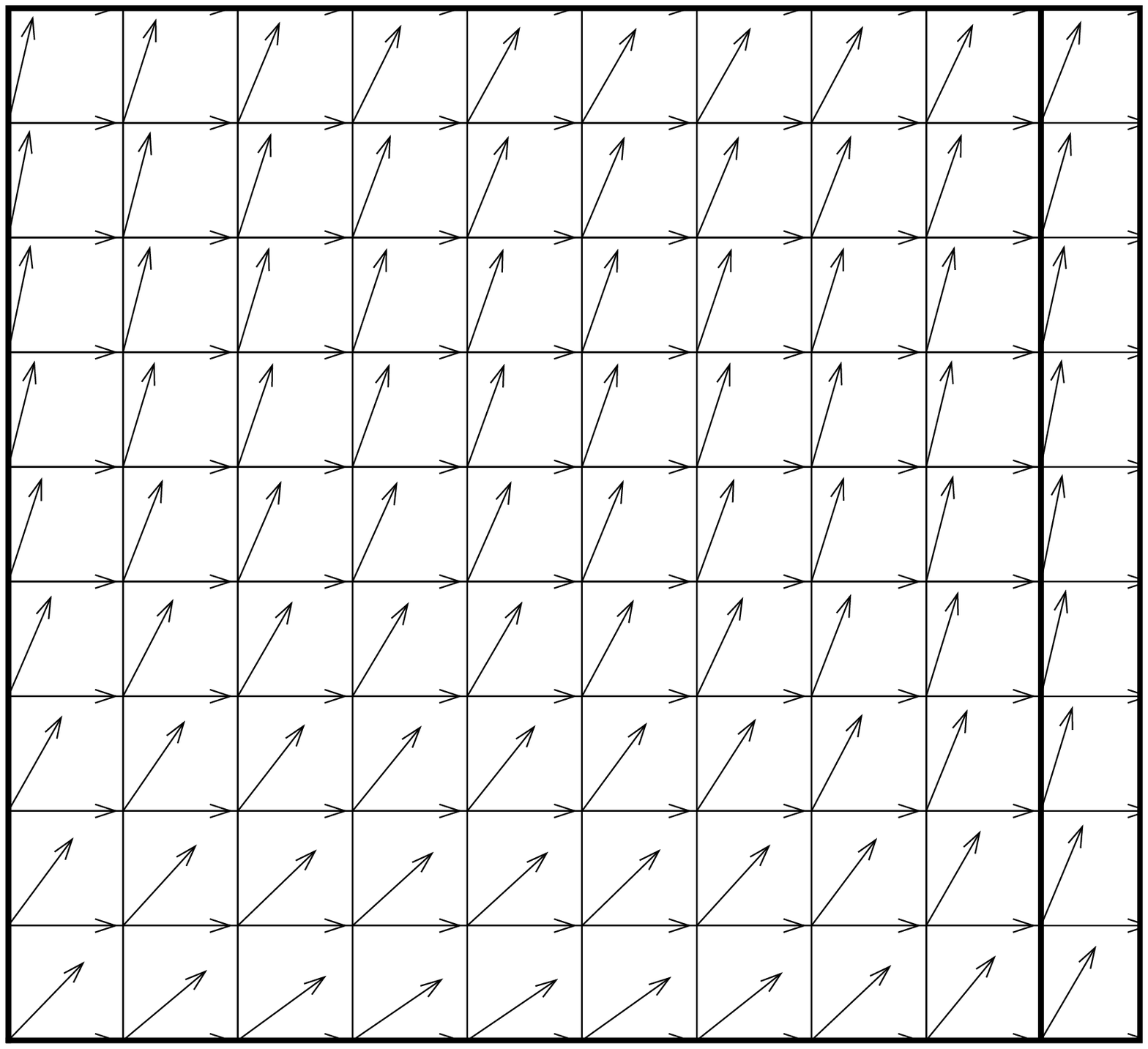}}}
  \caption{The two metastable states exhibited by the spin valve system at room temperature: a 'C' state and a 'S' state. The horizontal arrows are those of the magnetization of the bottom layer which points along the x-axis.}
\label{config}
\end{figure}

The application of stochastic resonance \cite{benzi} to increase the signal 
to noise ratio (SNR) is now well established in many physical
systems \cite{sr} and has also been shown
to increase the stability of an unstable system in some 
cases \cite{dayan}. \ Particle production in a quantum field can
also be 
enhanced in the presence of
 noise \cite{ishihara}. \ The SNR is enhanced 
by adding noise to the system 
as it is being driven by an external periodic force.  \ However 
until now only one-dimensional systems with additive noise and 
a simple bistable  potential, e.g., $U(x) = x^4-x^2$, or 
a potential  with one metastable
state, e.g., $U(x)=x^3-x^2$,
 have been 
studied and shown to have this interesting property of increasing 
the SNR with the addition of noise  as it is being driven 
by a time-dependent  force. \ This counter intuitive 
response is inherently related to the non-linearities present in
 these 
systems. \ One might therefore ask if a 
similar behavior can be found in more 
complicated systems, in particular magnetic 
systems. \ In this letter, we give one example of 
such systems; a spin valve of two   magnetic layers separated
by a normal conductor. \ The magnetization is 
 inhomogeneous  in the
'free' layer 
and a 
current perpendicular to the plane (CPP) traverses 
the structure. \ In this case a  
spin torque between the layers can exist and will play in the 
following an important role in the SR effect.

\ Until recently, the ideas of stochastic resonance have not been 
exploited in magnetic systems.\ As far as I know, 
reference \cite{marco} provides 
the only application 
we are aware of; it applies SR to the measurement of 
hysteresis loops in magnetic systems described by Preisach 
model. \ In 
the following, we study a  magnetic system 
with multiplicative noise and show that a 
stochastic resonance type behavior
exist in this system and can be taken advantage of to {\it selectively}
suppress frequencies from the noise spectrum. \ The system we treat 
is very realistic and hence more complex than 
previously treated systems \cite{sr}.\ The use of strong time-dependent 
currents implies that our system is in a non-equilibrium state and 
linear response methods are 
obsolete for our study \cite{pankratov}. \ In this 
letter we will attempt  
only a qualitative treatment of this subject aided by numerical 
integration
 of the non-linear Landau-Lifshitz-Gilbert equation \cite{brown}.  

\ We study a current perpendicular
to the plane (CPP) spin valve with two 
magnetic layers (fig. \ref{geometry}), one with 
 magnetization $\mathbf{S}_p$ pinned along the x-axis 
(reference layer RL)
 while the other  layer (FL) has a free 
magnetization $\mathbf{S}_f$. \ Figure \ref{ZeroCurrent} shows the 
power spectral densities (PSD) of the thre components of the 
magnetization in the absence of current. \ The PSD's have 
been normalized the same way in all the figures. \ The noise in 
the z component is smallest due to the demagnetization 
field.\ The current has an ac component in 
addition to the dc component. \ The source of noise 
in the system is due to thermal 
fluctuations that activates switching between two states 
 and hence it is intrinsic to the 
system. \ To get an energy surface with more than one local 
minimum,  we bias the spin valve with an external field that is 
close to being perpendicular 
to the fixed magnetization of the RL and take into 
account spin momentum transfer effects \cite{slon} 
between the two
layers of the spin valve. \ The 
y-component of the external field
 is kept at $600 \; Oe$ while 
the one along $\mathbf{S}_{p}$ is kept around $-100 \; Oe$. \ The 
layers have dimensions $100 \times 100 \times 3 \; nm^3$ with the 
easy axis along the x-axis and anisotropy $H_k = 50 \; Oe$. \ The 
magnetization
in the pinned layer is fixed by a large bias field. \ Hence, in 
this study the pinned layer is studied micromagnetically on equal 
footing with the free layer. \ The 
magnetization in this structure shows two configurations  
(Fig. \ref{config})
 which are non-homogeneous and are a result of the demagnetization
field, the current field and any interaction between the layers
 \cite{rebei}. \ In addition to stable and unstable 
states, the system has 
saddle points which are present due to the spin torque term. \ Spin 
valve 
structures are useful components of giant magneto-resistance (GMR)
sensor devices and hence 
the manifestations of stochastic resonance in these systems may be of 
practical importance. \ To increase the sensitivity of a GMR sensor, the bias 
field on the free layer is almost perpendicular to the magnetization 
of the pinned layer. \ It is this kind of biasing that gives rise to 
the $1/f$-type noise studied here since 
it permits the system to hop between 
 two-states. \ The results reported 
here are valid even if the pinning of the bottom layer 
is not perfect. \ In fact, large bias fields tend to 'distort' the 
magnetization in the pinned layer but only slightly and will 
not contribute to the 1/f-type noise observed here.

\ To analyze 
this system, we use the Landau-Lifshitz-Gilbert equation \cite{brown} 
supplemented by the spin torque as first suggested by Slonczweski 
 \cite{slon},

\begin{eqnarray}
\frac{d \; \mathbf{M}}{d\; t} &=& -|\gamma| \mathbf{M}\times 
\left(
 \mathbf{H}_{eff} +  \mathbf{h} (t)  \right. \\
&& \left. + p \frac{I_s}{M_s} \left(\mathbf{M}\times \mathbf{m}_{ref} \right)- \frac{\alpha}{|\gamma| M_s} \frac{d\; \mathbf{M}}{d\; t} \right) \nonumber
\end{eqnarray}
where $\gamma$ is the gyromagnetic ratio and $\alpha$ is the damping
constant taken here to be equal to 0.01. \ The 
effective field $\mathbf{H}_{eff}$ includes exchange 
interactions, anisotropy, the demagnetization field
 and the 
field from the current.\ The interlayer 
exchange is assumed small of the order of 
$ 20 \; Oe$. \ The value of $\alpha$ chosen is 
larger than typical values of damping in the bulk of a Permalloy. \ In a 
geometry such as ours, the damping is dominated by interfacial 
damping and 0.01 is a reasonable value for it \cite{rebei2}.
\ The noise is assumed
 white Gaussian, $
\langle h_i(t)h_j(t^\prime) \rangle = 2 D \delta_{ij} \delta(t-t^\prime)$ with $i,j=1,2,3$ 
and using the 
fluctuation-dissipation theorem (FDT), $D$ is proportional to the 
temperature $T$ and inversely proportional to the volume $V$,
 $D = 2 k_B T/(|\gamma|M_s V)$ \cite{zhang}.  \ However, as we have 
shown in \cite{rebei2}, in the presence of a bias voltage the FDT 
 is strictly not applicable unless we 
restrict our region of interest to
frequencies close 
to the FMR frequency of the system. \ In the following, we assume that
FDT is still applicable. \ $\mathbf{m}_{ref}$ 
is the direction of the magnetization in the 
RL. \ The current $I_s = I + i\; sin\left(\Omega t \right)$ has 
 one static
 component and another one with frequency $\Omega$. \ The form of 
the spin torque in the presence of ac currents 
still has the same form as in the Slonczewski 
equation. \ Zhu et al. \cite{zhu} 
calculated the ac spin torque in a magnetic tunnel junctions and found 
it to be very close in form to the static case. \ Among other things, 
they found that at low frequencies, the total spin torque can become 
less than the spin torque of the static 
component of the current. \ Hence an ac current can introduce 
a partial self-cancellation of the spin torque. \ Therefore
we expect the ac spin torque to affect the 
noise spectrum 
of a magnetic system in a nontrivial way. \
\ The 
noise or spectral density functions of the system are calculated numerically 
since for time dependent systems the fluctuation dissipation relations are 
invalid \cite{rebei2}.\ The current flows along the z-direction perpendicular 
to the xy plane of the magnetic layers. \ The magnetization of the 
RL is fixed 
along the x-direction.\ The FMR frequency for this system is around 
$4.5 \; GHz$ and $M_s = 1400 \; emu/cc$.\ The 
constant $p = 100 \; Oe/mA$. \ In magnetic recording,e.g., such 
devices operate at 
frequencies below the FMR frequency and hence this system is 
considered noisy to be useful as a 
sensor. \ One possible way to address this problem is to 
seek a way to suppress the noise for frequencies less than 
$ 1.0 \; GHz$ and may be shift the noise to higher frequencies which 
is outside the operational range of the device. \ In this sense we are 
selectively suppressing the noise below the FMR 
frequency only. \ Previous applications of 
stochastic resonance were interested in suppressing the noise around the 
frequency of the driving force which is usually the signal to be 
measured.

\begin{figure}[htbp]
\centering{\resizebox{4cm}{!}{\includegraphics{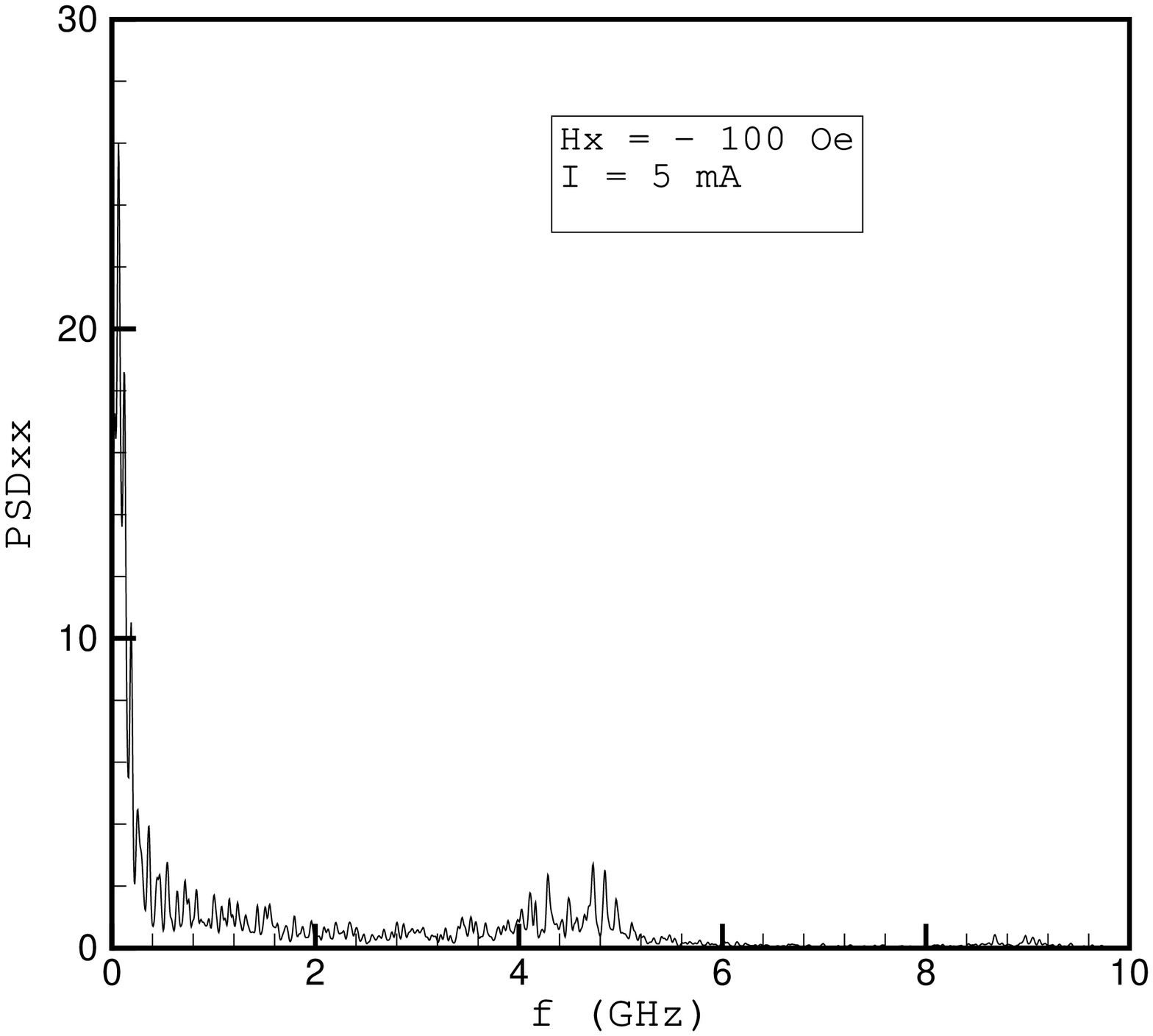}}}
\centering{\resizebox{4cm}{!}{\includegraphics{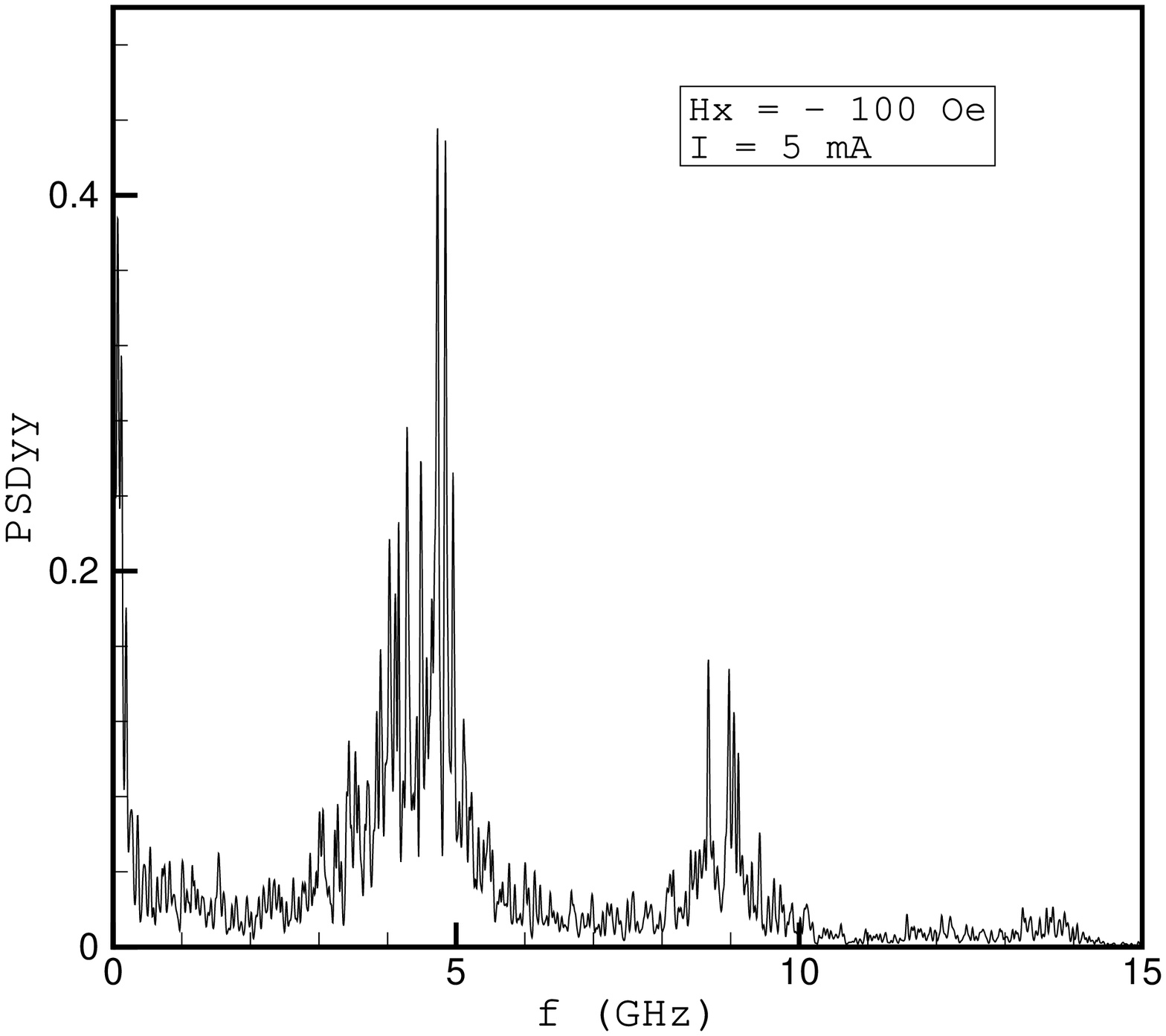}}}
\centering{\resizebox{4cm}{!}{\includegraphics{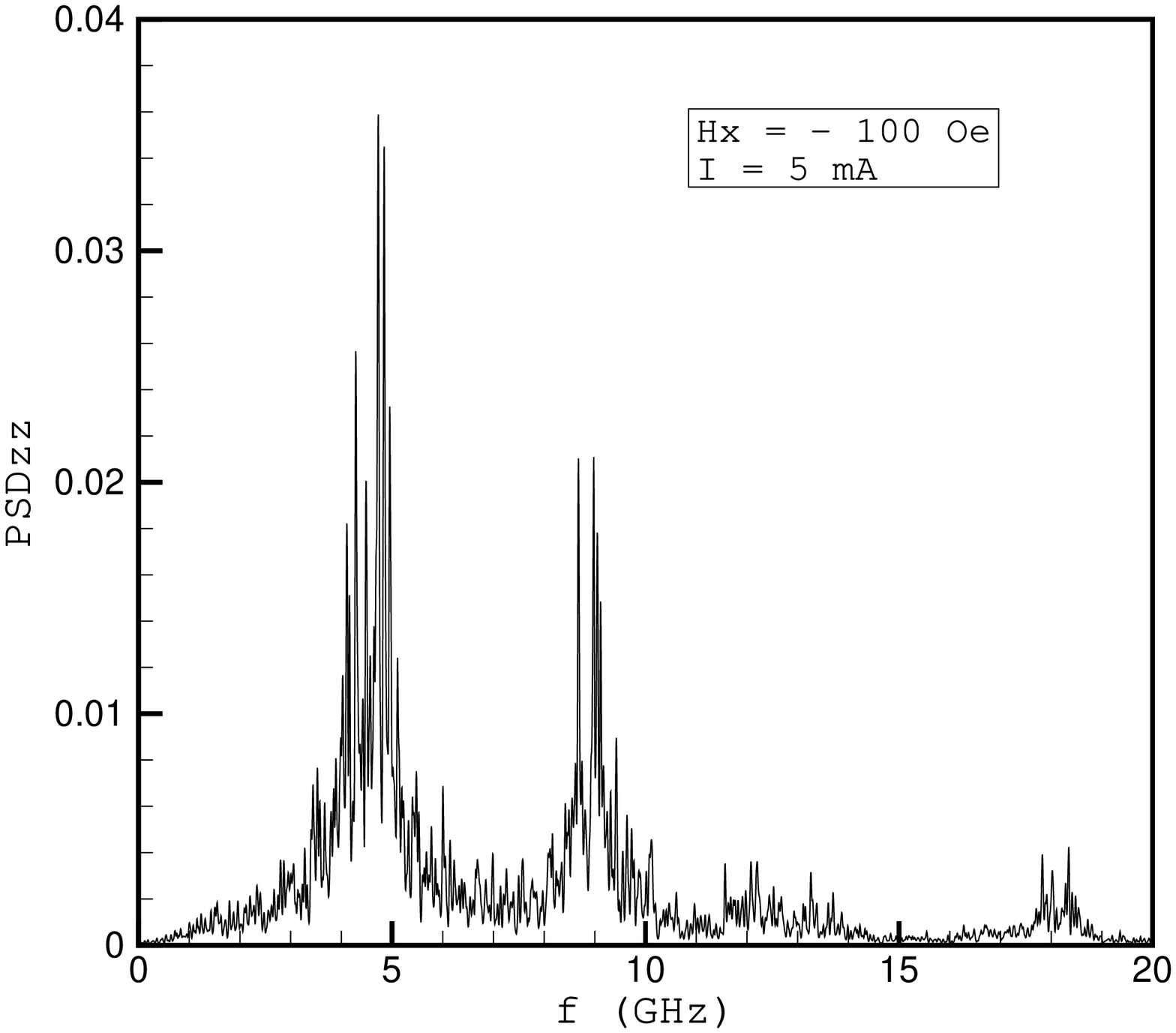}}}
  \caption{{{The noise spectrum in the x-component of the magnetization of the FL in the absence of ac current: 1) x-component 2) y-component 3) z-component. The x-component component of the magnetization shows significant low frequency noise compared to the other components. The major two peaks in the y and z components are due to the inhomogeneous two states in the system.}}}
\label{xac0}
\end{figure}
\begin{figure}[htbp]
\centering{\resizebox{4cm}{!}{\includegraphics{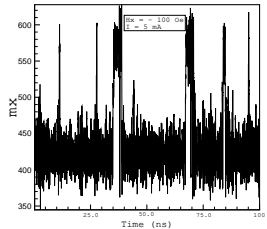}}}
  \caption{{{The x-component of the magnetization as a function of time. The behavior of the average x-component indicates that the magnetization is switching between two states, one stable and the other is unstable (a saddle point). }}}
\label{xtac0}
\end{figure}
\begin{figure}[htbp]
\centering{\resizebox{4cm}{!}{\includegraphics{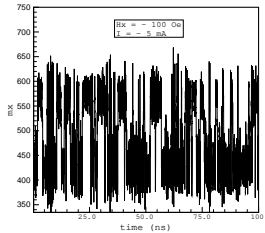}}}
  \caption{{{The x-component of the magnetization as a function of time for opposite sign of the dc current in fig. \ref{xtac0}. The spin torque appears in this case to change the topology of the energy surface to one where both states are equally visited by the magnetization.}}}
\label{xt5nac0}
\end{figure}


\ From Fig.\ref{xac0}, we observe
 that it is the x-component of the magnetization 
of the FL that is the noisiest and hence this 
would interfere with the GMR signal. \ The out-of plane z-component of the 
magnetization is very quiet due to the demagnetization 
field. \ Fig. \ref{xtac0} shows that the source of the 
noise is from the switching of the magnetization between two configurations. \ One 
with an average x-component of about $430 \; emu/cc$ and the other less 
stable one with an average x-component around $580 \; emu/cc$.\ The spectral
noise in the y and z components show more than the usual FMR peak 
since the system is not in equilibrium. \ The 
higher order harmonics are due to non-homogeneity of the 
magnetization, i.e., spin waves.

\ In fig. 
\ref{xt5nac0}, we change the sign of the current to show the effect
of the spin torque on the magnetization and noise. \ In this case, 
the magnetization spends almost equal time in both states. \ This suggests 
the use of the spin torque itself as a regulator of the transition rate
between two bistable states.\ In this letter, we will not pursue this idea 
further and instead we want to explore the idea of adding a strongly 
periodic current in addition to the dc bias current in order to suppress 
the noise for frequencies below $ 1.0 \; GHz$.

\begin{figure}[htbp]
\centering{\resizebox{4cm}{!}{\includegraphics{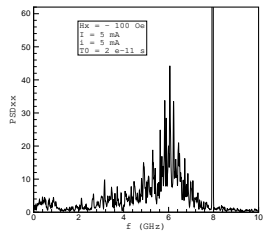}}}
  \caption{{{The noise spectrum in the x-component of the magnetization of the FL in the presence of an ac component to the current with same amplitude as the dc part. The low frequency part of the 
spectrum has been completely suppressed by the addition of the ac current. The peak at $8 \; GHz$ is that of the ac current.}}}
\label{xac5}
\end{figure}

\begin{figure}[htbp]
\centering{\resizebox{4cm}{!}{\includegraphics{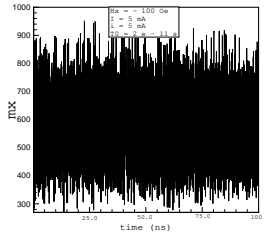}}}
  \caption{{{The noise spectrum in the x-component as a function of time in the presence of the ac current. The energy surface in this case appears to have only one stable minimum and no switching is observed. The noise in the x component has been pushed to high frequencies.}}}
\label{xac5t}
\end{figure}

\ Figure \ref{xac5} shows the result of applying an ac current
with amplitude equal 
to the bias current and a frequency of about $8 \; GHz$ for the same 
system of fig. \ref{xac0}. \ The noise spectrum for frequencies below 
$ 1 \; GHz$ is greatly suppressed by this additional current 
source. \ However now the peak  around the FMR frequency is much more
pronounced and so does its width. \ The second narrow peak is that 
of the ac current. \ A closer look at the real time behavior of the 
x-component of the magnetization, fig. \ref{xac5t}, implies that 
the original energy barrier is no longer present. \ The strong 
periodic
current which in turn induces
 a periodic spin torque is driving the system and 
making it less sensitive to random 
thermal fluctuations which are the source of the low frequency. \ To 
be an effective suppressant, the frequency of the torque has to be 
outside the bandwidth of the sensor device.

\begin{figure}[htbp]
  \centering{\resizebox{4cm}{!}{\includegraphics{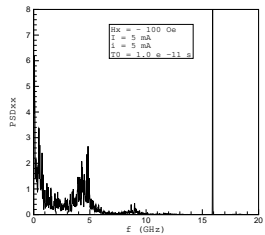}}}
  \caption{{{The noise spectrum in the x-component of the magnetization of the FL. \ The frequency of the ac current has been doubled compared to fig. \ref{xac5}.}}}
\label{xac0hT1}
\end{figure}

\  Finally, fig. \ref{xac0hT1} shows that an increase in the frequency
of the ac current degrades the effectiveness of the ac current component to suppress the low frequency noise. \ We found for this example, 
that for current amplitudes at $ 5 \; mA$, the maximum frequency the 
current should have is about twice the FMR frequency of the system. \ This latter criterion seems to depend strongly on the energy surface and
hence there is no  universal behavior 
as in the bistable one dimensional 
well where the frequency of the driving force is the inverse of 
twice the inter-well transition time.

\ In summary, we have shown that the ideas of SR are 
still attractive even for complex systems which can not be described 
by a one dimensional potential. \ In higher dimensions and 
in the presence of a spin torque, saddle point 
states become available to the system and may invalidate the 
ideas of stochastic resonance. \ In this letter, we showed 
that a strongly periodic current can suppress thermally induced 
jumps between a stable and a metastable state and hence 
suppresses the low frequency noise and enhance the FMR peak in the 
spin valve. 

\bigskip

{\it This work is based on work done in the Fall of 2004 at Seagate Research and will not be published elsewhere. It is merely a compilation of numerical 
observations. 
The author hopes to convince others that 
there is an interesting interplay of noise 
and time-dependent spin torques in CPP spin valves that is worth exploring in a more systematic way than attempted here.}

\section*{Acknowledgements}

We would like to thank G.J. Parker for making
 his LLG solver available to us.

\end{document}